\documentclass{PoS}
\usepackage{bm}
\usepackage{amssymb}
\usepackage{amsmath}
\usepackage{amsfonts}
\usepackage{dsfont}
\usepackage{subfigure}
\usepackage{color}

\newcommand{\ket}[1]{\left| #1 \right>}

\title{Gaussian states for the variational study of (1+1)-dimensional lattice gauge models}

\ShortTitle{Gaussian states for the variational study of (1+1)d LGT}

\author{P. Sala\\
       Department of Physics, T42, Technische Universit{\"a}t M{\"u}nchen, James-Franck-Stra{\ss}e 1, D-85748 Garching, Germany\\
       E-mail: \email{pablo.sala@tum.de}}
       
\author{T. Shi\\
       Institute of Theoretical Physics, Chinese Academy of Sciences, P.O. Box 2735, Beijing 100190, China\\
       E-mail: \email{tshi@itp.ac.cn}}
       
\author{\speaker{S. K{\"u}hn}\\
       Perimeter Institute for Theoretical Physics, 31 Caroline Street North, Waterloo, ON N2L 2Y5, Canada\\
       E-mail: \email{skuhn@perimeterinstitute.ca}}
       
\author{M. C. Ba{\~n}uls\\
       Max-Planck-Institut f\"ur Quantenoptik, Hans-Kopfermann-Stra{\ss}e 1, 85748 Garching, Germany\\
       E-mail: \email{banulsm@mpq.mpg.de}}
       
\author{E. Demler\\
       Department of Physics, Harvard University, Cambridge, Massachusetts 02138, USA\\
       E-mail: \email{demler@physics.harvard.edu}}
       
\author{J. I. Cirac\\
       Max-Planck-Institut f\"ur Quantenoptik, Hans-Kopfermann-Stra{\ss}e 1, 85748 Garching, Germany\\
       E-mail: \email{ignacio.cirac@mpq.mpg.de}}

\abstract{We introduce a variational ansatz based on Gaussian states for (1+1)-dimensional lattice gauge models. To this end we identify a set of unitary transformations which decouple the gauge degrees of freedom from the matter fields. Using our ansatz, we study static aspects as well as real-time dynamics of string breaking in two (1+1)-dimensional theories, namely QED and two-color QCD. We show that our ansatz captures the relevant features and is in excellent agreement with data from numerical calculations with tensor networks.}

\FullConference{The 36th Annual International Symposium on Lattice Field Theory - LATTICE2018\\
		22-28 July, 2018\\
		Michigan State University, East Lansing, Michigan, USA.}

\begin{document}

\section{Introduction}
\vspace{-0.75em}
Originally pioneered by Wilson~\cite{Wilson1974}, lattice gauge theory (LGT) has become a central tool for exploring gauge models in the nonperturbative limit. In particular, the discretization of the action on a Eulidean spacetime lattice allows for a numerical approach with Monte Carlo (MC) methods which has made it possible to gain major insights into the mass spectrum of QCD~\cite{Durr2008}, string breaking~\cite{Bali2005} and many other phenomena. However, in certain parameter regimes the onset of the sign problem~\cite{Troyer2005} prevents an efficient MC sampling. As a result certain questions, such as the phase diagram of QCD at finite temperature and nonvanishing baryon chemical potential, are intractable with the method. Moreover, real-time dynamics are mostly inaccessible.

In recent years, variational approaches to the Hamiltonian lattice formulation~\cite{Kogut1975} have proven themselves as promising alternatives. For one, methods based on tensor networks~\cite{Dalmonte2016} have been applied to various (1+1)-dimensional Abelian and non-Abelian gauge models. Using matrix product states (MPS), a particular kind of one-dimensional tensor network, the major promise of overcoming the sign problem~\cite{Banuls2016a,Silvi2016} as well as simulations of real-time dynamics~\cite{Pichler2015,Kuehn2015,Buyens2016b} have already been successfully demonstrated. However, simulation of out-of-equilibrium dynamics and the generalization to higher dimensions remain computationally challenging~\cite{Zapp2017}.

Here we introduce a different variational ansatz for LGTs in 1+1 dimension based on Gaussian states~\cite{Bravyi2005,Kraus2009,Peschel2009}, meaning states whose density matrix can be expressed as the exponential of a quadratic form of the creation and annihilation operators. To this end we derive a set of unitary transformations which decouple the gauge degrees of freedom (d.o.f.) from the fermionic matter and turn the Gaussian states (in the correct frame) into suitable variational ansatzes for gauge models with fermionic matter. In the following we review the approach (Sec. \ref{sec:model_methods}) and some of our results for (1+1)-dimensional QED (Sec. \ref{sec:res_U1}) and two-color QCD (Sec. \ref{sec:res_SU2}) from Ref.~\cite{Sala2018}.

\section{Model \& Methods\label{sec:model_methods}}
We study the Kogut-Susskind lattice Hamiltonian with  staggered fermions. For a system with open boundary conditions on a lattice with $N$ sites the Hamiltonian reads~\cite{Kogut1975}
\begin{align}
H =\varepsilon\sum_{n=1}^{N-1}\left( {\bm{\phi}_{n}^{\dagger }U_{n} \bm{\phi}_{n+1}+\mathrm{H.c.}}\right) + m \sum_{n=1}^N{(-1)^{n}\bm{\phi}_{n}^{\dagger }\bm{\phi}_{n}}+\frac{g^2}{2}\sum_{n=1}^{N-1}{\bm{L}_{n}^{2}}.
\label{eq:KS_Hamiltonian}
\end{align}
Here $U_n$ is a group element in the fundamental representation acting on the gauge links, and the fermionic field $\bm{\phi}_{n}$ is a spinor in the same representation acting on the matter residing on the sites. Eq. \eqref{eq:KS_Hamiltonian} can be seen as the lattice discretization of a continuum model, if we set $\varepsilon=1/2a$, $m=m_0$, $g=g_0^2a$, where $m_0$ and $g_0$ are the bare mass and coupling of the model and $a$ denotes the lattice spacing. Physical states, $\ket{\psi}$, have to fulfill Gauss law $G_{n}^{a}\ket{\psi}=0$ $\forall a,\,n$, where 
\begin{align*}
G_{n}^{a}=L_{n}^{a}-R_{n-1}^{a}-\mathcal{Q}_{n}^{a}
\end{align*}
are the generators of time-independent gauge transformations. In the expression above $L_{n}^{a}$ ($R_{n}^{a}$) denote the components of the left (right) color-electric field on a link and $\mathcal{Q}_{n}^{a}$ the components of the total charge which is the sum of the dynamical charges $Q^a_n$ and the (static) external charges $q^a_n$.

In the following, we are interested in two particular cases, (1+1)-dimensional QED also known as the Schwinger model, for which the gauge group is U(1), and the non-Abelian case of two-color one-dimensional QCD with gauge group SU(2). For the Schwinger model $\bm{\phi}_{n}$ is simply a single component fermionic field, $\phi_{n}$, and $\bm{R}_{n}=\bm{L}_{n}=L_{n}$ because of the Abelian nature of the gauge group. The (staggered) charge is given by $Q_{n}=\phi _{n}^{\dagger }\phi_{n}-(1-(-1)^{n})/2$ and external charges are just real numbers, $q_{n}\in\mathds{R}$. In the strong-coupling limit, $g^2/\varepsilon\gg 1$, the hopping term in Eq. \eqref{eq:KS_Hamiltonian} can be neglected and the Hamiltonian can be solved analytically. The gauge invariant ground state in the sector of vanishing total charge is given by the lattice analog of the Dirac sea which reads
\begin{align}
|\psi_{\text{SC},\text{U(1)}}\rangle = |\mathbf{1};\mathbf{0};\mathbf{1};\mathbf{0};\dots \rangle \otimes |0\rangle_\text{gauge}.
\label{eq:SC_U1}
\end{align}
In the expression above the numbers in bold face indicate the occupation of the fermionic sites and $|0\rangle_\text{gauge}$ the gauge links carrying no flux. 

For the case of SU(2) we have two colors of fermions (``red'' and ``green'') and, hence, the fermionic fields $\bm{\phi}_{n}$ have two components, $\bm{\phi}_{n}=(\phi _{n}^{r},\phi _{n}^{g})^{T}$, taking into account the ``red'' (${\phi}_{n}^r$) and ``green'' (${\phi}_{n}^g$) fermions. The total SU(2) color charge reads $\bm{\mathcal{Q}}_{n}=\bm{Q}_n+\bm{q}_n$ where the components of the dynamical charges are given by $
{Q}_{n}^{a}=\frac{1}{2}\bm{\phi}_{n}^{\dagger }\sigma ^{a }\bm{\phi}%
_{n}$, $a=x,y,z$, and $\sigma^a$ are the usual Pauli matrices. For a nonvashing external charge at site $n$, $q_{n}^{a}=\frac{1}{2}\sigma ^{a}$, otherwise it is zero. Analogously to the case of U(1), the strong-coupling limit can be solved analytically yielding
\begin{align*}
 |\psi_{\text{SC},\text{SU(2)}}\rangle = |\mathbf{1,1};\,\mathbf{0,0};\,\mathbf{1,1}\dots\rangle \otimes |0\rangle_\text{gauge}
\end{align*}
for the ground state. Again the numbers in bold face indicate the fermionic occupation for the two colors on a site and $|0\rangle_\text{gauge}$ the gauge links carrying no flux.

Since there are no transversal d.o.f. in the (1+1)-dimensional case, it is possible to disentangle the gauge and matter d.o.f. completely with a unitary transformation. For our lattice system with open boundary conditions this transformation is given by\footnote{For a detailed derivation see Ref.~\cite{Sala2018}.} 
\begin{align*}
\Theta =\prod_{k=1}^{\rightarrow }{\exp \Big(i \bm{\theta}_{k}\cdot
\sum_{m>k}{\bm{\mathcal{Q}}_{m}\Big)}},
\end{align*}
where the arrow indicates an ordering of the factors in the product according to increasing $k$. In the sector of vanishing total charge applying this transformation to the Hamiltonian yields
\begin{align}
  H_{\Theta }=\varepsilon \sum_{n}\left( {\bm{\phi}_{n}^{\dagger } \bm{\phi}_{n+1}+}\text{H.c.}\right)  
 +m \sum_{n}{(-1)^{n}\bm{\phi}_{n}^{\dagger }\bm{\phi}_{n}}
 +\sum_{a}\sum_{n,m}\mathcal{Q}_n^aV_{n,m}\mathcal{Q}_m^a
\label{eq:H_Theta}
\end{align}
with $V_{n,m} = -\frac{1}{2}|n-m|$. Contrary to the original Hamiltonian, $H_\Theta$ only depends on the fermionic d.o.f. and exhibits long-range interactions between the charges in the color-electric term. 

Our goal is to solve the Hamiltonian \eqref{eq:KS_Hamiltonian}. To this end we would like to compute the evolution of an initial state $\ket{\psi}$ in either imaginary or real time to obtain the ground state or simulate the dynamics. Following Ref.~\cite{Shi2018}, we choose our variational ansatz to be
\begin{align}
\ket{\psi}=\Theta^{\dagger}U_{\text{ext}}\ket{\text{GS}} { \left\vert 0\right\rangle _{\text{gauge}}}
\label{eq:Gaussian_ansatz}
\end{align}
where $\ket{\text{GS}}$ is a fermionic Gaussian state and $U_{\text{ext}}$ another unitary transformation which decouples the external from the dynamical charges (see Sec. \ref{sec:res_SU2}). Notice that this ansatz is non-Gaussian, since the transformations $\Theta$, $U_\text{ext}$ are in general not an exponential of a quadratic form of the fermionic and bosonic operators. Instead of studying the dynamics of $\ket{\psi}$ under the Hamiltonian, we can equivalently compute the evolution of $\ket{\text{GS}}$ under $H_2=U_{\text{ext}}H_\Theta U_{\text{ext}}^\dagger$. The
time-dependent variational principle within the family of Gaussian states is implemented via the evolution equations 
\begin{align}
\frac{d}{d\tau }\Gamma (\tau )=\{\Gamma ,\mathcal{H}(\Gamma )\}-2\Gamma
\mathcal{\ H}(\Gamma )\Gamma,\quad\quad i\frac{d}{dt}\Gamma (t) =[\mathcal{H}(\Gamma ),\Gamma ],  \label{eq:EOM}
\end{align}
for imaginary time $\tau$ and real time $t$, as shown in Ref.~\cite{Shi2018}. Here $\Gamma$ is the covariance matrix collecting all two-point correlation functions of the Gaussian state, and $\mathcal{H}$ an (effective) single-particle Hamiltonian which can be computed efficiently. In the following we use Eq.~\eqref{eq:EOM} to study static and dynamical aspects of string breaking in the sector of vanishing total charge.

\section{U(1) Gauge Theory\label{sec:res_U1}}
\begin{figure}[htp!]
\centering
\includegraphics[width=0.245\textwidth]{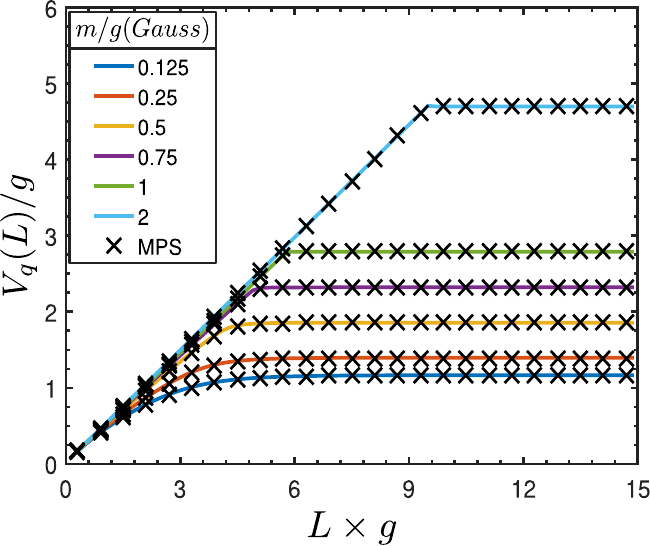}
\includegraphics[width=0.245\textwidth]{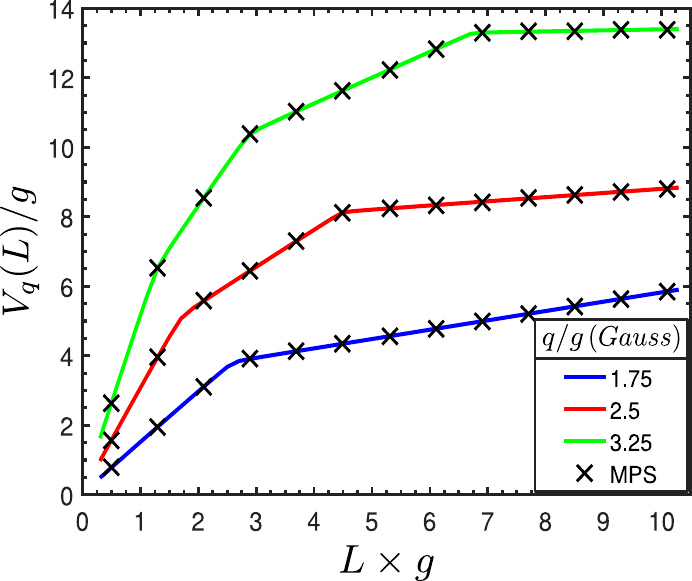}
\includegraphics[width=0.245\textwidth]{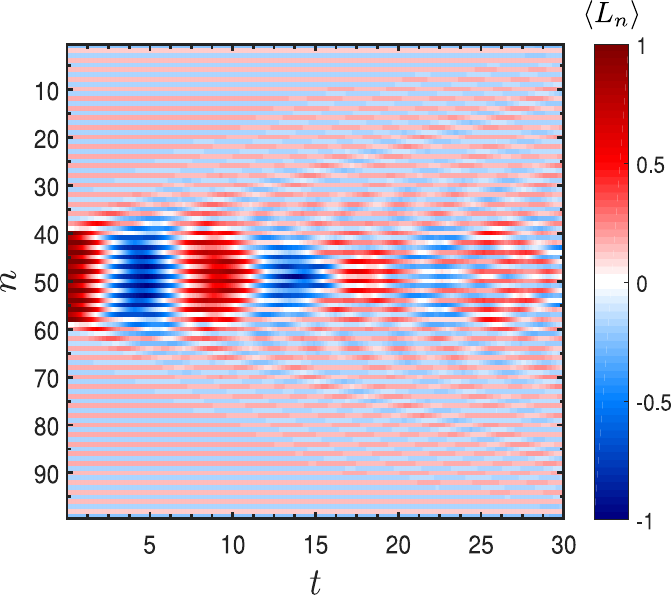}
\includegraphics[width=0.245\textwidth]{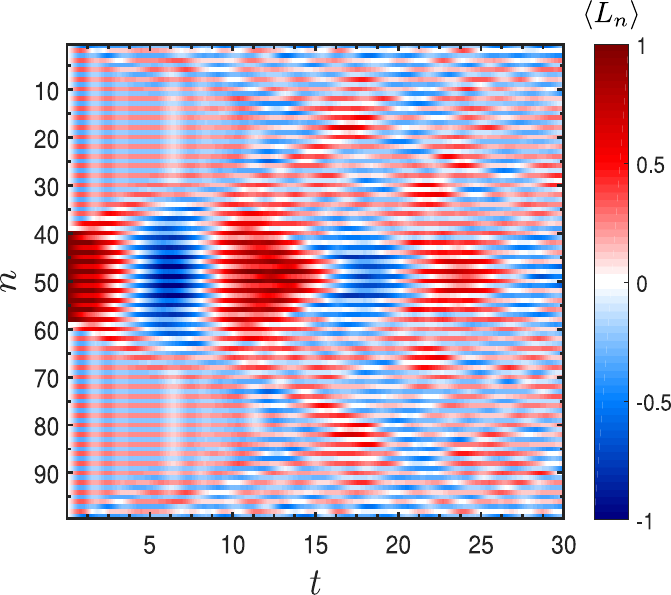}
\put(-340,78){{\small(a)}}
\put(-230,78){{\small(b)}}
\put(-138,78){{\small(c)}}
\put(-30,78){{\small(d)}}
\caption{(a) Static potential for integer external charges $q/g=1$, $1/(ag_0)^2=100$ and various masses. (b) Static potential for various noninteger external charges  $1/(ag_0)^2=100$ and $m/g=1$. The solid lines represent the results from our variational ansatz, the black crosses the results from the MPS calculation from Ref.~\cite{Buyens2015}. (c) Evolution of the electric flux distribution $\langle L_n\rangle$ for a string of length $L/a=19$ imposed on top of the interacting vacuum with $Q/g=1$, $m=0.1$ and $g=0.1$. (d) Same evolution as in (c), but with the string imposed on the strong-coupling vacuum.}
\label{fig:U1}
\end{figure}

\vspace{-0.5em}
Let us first turn to the Abelian case of (1+1)-dimensional QED. Since the external charges in that case are simply real numbers, $U_{\text{ext}} = \mathds{1}$. Using our variational ansatz, we can investigate the static potential between two external charges similar to conventional MC simulations~\cite{Bali2005}. First we determine the vacuum energy, $E_\text{vac}$, by evolving \eqref{eq:SC_U1} in imaginary time. Subsequently, we place a pair of external charges separated by a distance $L$ on top of the strong-coupling vacuum. As a consequence of Gauss law, they have to be connected by an electric flux tube. Again, we evolve the resulting state in imaginary time using Eq. \eqref{eq:EOM} to find its ground state energy $E_q(L)$. In Fig.~\ref{fig:U1}(a) we show our results for the static potential $V_q(L) = E_q(L)-E_\mathrm{vac}$ for external charges which are integer multiples of the coupling $g$. We clearly observe that for short distances the static potential grows linearly, thus indicating the presence of a flux tube in the ground state. From a certain length $L_c$ on the potential saturates, showing that the string is broken and particle-antiparticle pairs have been created which screen the external charges. Figure \ref{fig:U1}(b) shows our results for noninteger values $q/g$ of the external charge. Again we observe a linear increase up to a certain length, but contrary to the previous case, the potential does not saturate because for noninteger $q/g$ the particle-antiparticle pairs cannot completely screen the external charges. In both cases, our results are in excellent agreement with the MPS results from Ref.~\cite{Buyens2015}.

Our variational ansatz is not limited to static problems and allows us to study the real-time dynamics of string breaking, too. In the following we compute the evolution of a fully dynamical string, for which the charges at the end can also propagate. Figure \ref{fig:U1}(c) shows the results for a string imposed on top of the interacting vacuum of length $L/a=19$. We observe that the string is breaking due to the creation of particle-antiparticle pairs and the oscillation between string and antistring configurations eventually stops.
Using the Gaussian ansatz, we can also study a global quench scenario and evolve the strong-coupling vacuum with a string imposed on top of it, as shown in Fig.~\ref{fig:U1}(d). While the evolution for short time scales is relative similar to the previous case, wave fronts are emerging from the edges as a result of the global quench. At later times those interfere with the string dynamics, thus leading to notable differences.

\section{SU(2) Gauge Theory\label{sec:res_SU2}}
\vspace{-0.5em}
Our approach is also suitable for non-Abelian cases which we illustrate with two-color QCD. While the vacuum energy in absence of external charges can be computed analogously to the Schwinger case, special care has to be taken in the presence of a pair of external charges. Since these are now described by spin operators, the physical states of $H_\Theta$ in the presence of a pair of external charges are not Gaussian. However, it is possible to decouple the external charges from the dynamical ones. Similar to Ref.~\cite{Ashida2018}, we can identify parity symmetries of $H_\Theta$, which allow us to find (non-Gaussian) unitary transformations $V_1$ and $V_2$ to decouple the external charges using $U_\text{ext}=V_2V_1$~\cite{Sala2018}. In the rotated frame, the external charges become classical variables and the Hamiltonian reads $H_2(s_1,s_2) = U_\text{ext}H_\Theta U_\text{ext}^\dagger$, where $s_1,s_2\in\{-1,1\}$ are related to the eigenvalues of the parity operators. Moreover, we can characterize the entanglement between two external charges at positions $n_1$ and $n_2$ using the gauge invariant correlation function
\begin{align*}
C_{2}(n_{1},n_{2})=\sum_{a,b }{\left\langle q_{n_{1}}^{a } \Big(U_{n_{1}}^{\text{Adj.},\dagger }\cdots U_{n_{2}-1}^{\text{Adj.},\dagger }\Big)_{a,b}q_{n_{2}}^{b}\right\rangle }  
\end{align*}
where $ U_{n}^{\text{Adj.}}$ is a group element in the adjoint representation acting on link $n$. In the rotated frame this translates to $C_{2}(n_{1},n_{2})=\sum_{a}\left\langle q_{n_{1}}^{a}q_{n_{2}}^{a}\right\rangle$.

\begin{figure}[htp!]
\centering 
\includegraphics[width=0.31\textwidth]{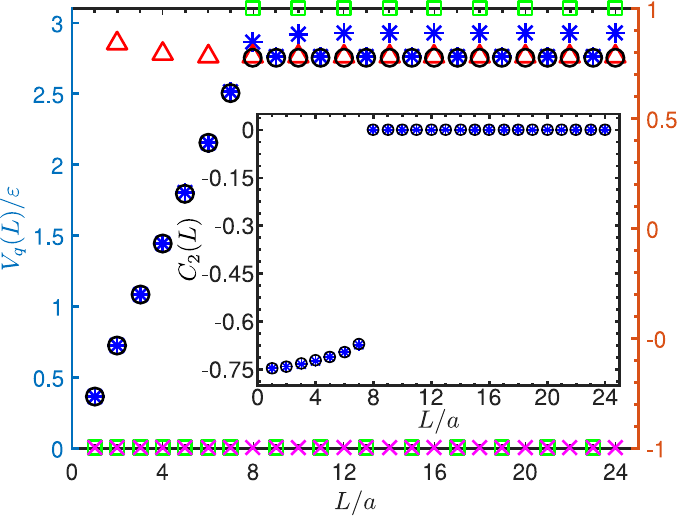} 
\includegraphics[width=0.35\textwidth]{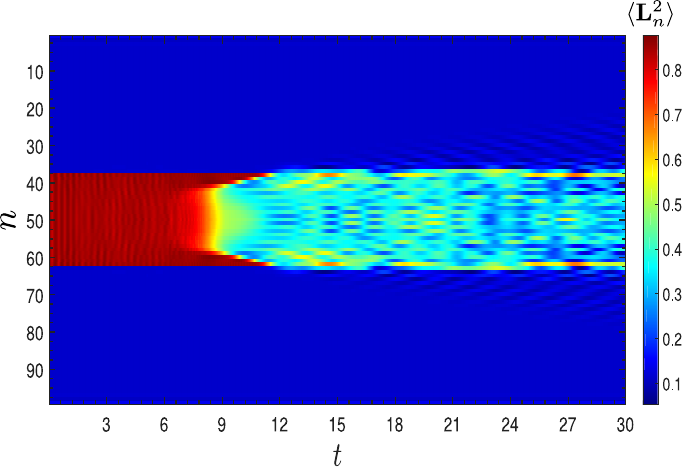}
\includegraphics[width=0.31\textwidth]{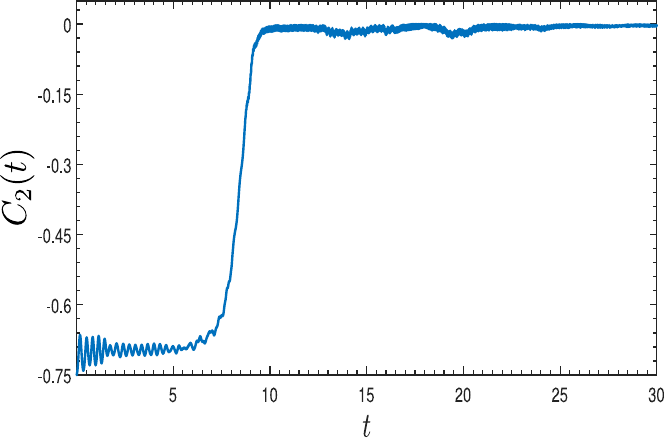}
\caption{(a) Static potential for $\varepsilon=1$, $m=0.5$, $g=1$ in the sectors $s_1=s_2=-1$ (blue asterisks) and $-s_1=s_2=-1$ (red triangles) as well as the MPS results (black circles). The right $y$-axis shows the expectation values for $s_1$ (green squares) and $s_2$ (magenta crosses) computed from the MPS ground state. Inset: Correlation function for the global ground state. (c) Spatially resolved color-flux profile as a function of time for a string between static external charges with length $L/a=25$ and $m=0.75$, $g=1.5$. (d) Corresponding correlation function between the external charges.}
\label{fig:SU2}
\end{figure}

In Fig.~\ref{fig:SU2}(a) we show our results for the static potential. Again we observe that the static potential increases linearly up to a certain critical separation $L_c$ of the charges. While the ground state for $L < L_c$ is in the sector $s_1=s_2=-1$, we see that this holds no longer true for $L\geq L_c$, for which the ground state for even lengths is in the sector $-s_1=s_2=-1$. The correlation function $C_2(L)$ also shows clear signatures of string breaking, as can be seen in the inset of Fig.~\ref{fig:SU2}(a). While the external charges are correlated for $L<L_c$, for which they are connected by a color-flux tube, they form color singlets with the surrounding dynamical fermions for $L\geq L_c$, the string breaks and $C_2\approx 0$. In order to benchmark our results, we also solve the Hamiltonian \eqref{eq:H_Theta} directly using MPS, which is in excellent agreement with the results from our ansatz Eq. \eqref{eq:Gaussian_ansatz}.

Again, we can also use our variational ansatz to study the real-time dynamics. In Fig.~\ref{fig:SU2}(b), we show the evolution of the spatially resolved color-flux profile for a string of length $L/a=25$ between static external charges imposed on top of the interacting vacuum. Initially we clearly observe an interacting string which is reflected by a region of high color flux and $C_2(0) \approx -3/4$ (see Fig. \ref{fig:SU2}(c)). At later times particle-antiparticle pairs are produced and the string eventually breaks resulting in $C_2(t)\approx 0$.

\section{Discussion \& Outlook}
\vspace{-0.5em}
We have introduced a variational ansatz based on Gaussian states for (1+1)-dimensional LGTs. The key idea is to identify unitary transformations which decouple the gauge and matter d.o.f. Using our ansatz, we have studied static and dynamical aspects of string breaking for the Schwinger model as well as for a non-Abelian SU(2) LGT. In both cases, we observe that the Gaussian ansatz is able to capture the relevant features and allows for reliable calculations of static properties as well as for out-of-equilibrium scenarios.

The unitary transformation $\Theta$ and $U_\text{ext}$ are not limited to the cases studied here, $\Theta$ works for arbitrary gauge groups SU(N) and $U_\text{ext}$ for any spacetime dimension. Moreover, the transformed Hamiltonian $H_\Theta$ might be useful for various other approaches. On the one hand, it can directly be addressed with tensor network methods without having to truncate the gauge d.o.f. On the other hand, it extends the formulation implemented in a recent quantum simulation experiment for the Schwinger model~\cite{Martinez2016} to non-Abelian groups. Since in the rotated frame there are no gauge d.o.f. anymore, this might allow for simpler experimental realizations than previous proposals for non-Abelian gauge models.

\section*{Acknowledgments}
\vspace{-0.5em}
We thank Erez Zohar for enlightening discussions and the authors of Ref.~\cite{Buyens2015} for sharing their data. PS acknowledges financial support from ``la Caixa'' fellowship grant for post-graduate studies. TS acknowledges the Thousand-Youth-Talent Program of China. SK was supported by Perimeter Institute for Theoretical Physics. Research at Perimeter Institute is supported by the Government of Canada through the Department of Innovation, Science and Economic Development Canada and by the Province of Ontario through the Ministry of Research, Innovation and Science. ED acknowledges support from Harvard-MIT CUA, NSF Grant No. DMR-1308435, AFOSR Quantum Simulation MURI, AFOSR-MURI: Photonic Quantum Matter, award FA95501610323. JIC acknowledges ERC Advanced Grant QENOCOBA under the EU Horizon 2020 program (grant agreement 742102). JIC and MCB are partly funded by the QUANTERA project QTFLAG.
\bibliographystyle{JHEP}
\bibliography{Papers_PoS}
\end{document}